# Statistical Theory of Initiation of Explosives by Impact

Gan Ren[*], Yingzhe Liu, Weipeng Lai and Tao Yu

*Xi'an Modern Chemistry Research Institute, Xi'an, 710065, People's Republic of China*

ABSTRACT

When a given weight dropped onto an explosive charge, explosion or not is probabilistic for certain impact energy and the frequency of explosion is always increase with increasing impact energy. Based on experimental results and recently theoretical work, we propose that the hot spot formation is attributed to the activated molecules decomposition and the number of molecules initiation is proportional to the impact energy but not the dropped weight heating as the previous hot spot theory. A theoretical model based on two states model has been put forward for this phenomena. It is shown that the activated molecules to form a hot spot determine the probabilistic nature of initiation by impact. It is shown a good agreement tested with Hexogen (RDX) experimental impact data.

**Keywords:** impact sensitivity, impact energy, two states model, hot spot, activated molecule

---

[*] E-mail:renganzyl@sina.com

When a given weight dropped onto an explosive charge, the initiation of charge shows a probabilistic character. This phenomenon has been observed a long time ago. Although the probabilistic character shows much difference under different testing conditions, yet the probability is stable under certain condition. This character is very important for the safe application and developing new energetic materials. Despite the importance, existing studies of the phenomenon were done most from the prediction of the height $h_{50}$ and mechanism of initiation but not from its probabilistic nature.

For its easily amenable to measurement, $h_{50}$ is the most popular criterion to characterize the impact sensitivity, and the most schemes are proposed to predict the height $h_{50}$. Most published works in height $h_{50}$ prediction are empirical, recently some combined with quantum chemistry calculation. The common ground of these works are attributed the impact sensitivity to single reason, including oxygen balance[1], molecular electronegativity[2], electronic band gap[3], electronic shakeup promotion energies[4], $^{13}$C and $^{15}$N NMR chemical shifts[5], and X-NO2 length[6-7] etc. The common deficiency is that these schemes can be only applied to small specified set of data and often misleading using to large set of data[8-9]. The same problems are also existed for multivariable schemes. The current most successful methods are the advance quantitative structure property relationships and regression techniques beyond straightforward multilinear regime etc[10-12], both get a nice R2 as applied to large set of data. The main deficiency of the scheme is it's from numerical but not from physical. Current more physical approach[13-16] is based on the assumption decomposition rate determine step to be the propagation of the decomposition reaction to adjacent molecules, combined with the all X-NO$_2$ bond dissociation contribution to sensitivity, the theory borrowed the idea from impact sensitivity and give a good agreement with testing data for a large set of explosives, which show can be easily used to predict new energetic material molecules.

Two theories of initiation by impact have been proposed to explore the mechanism: the thermal and non-thermal theories[17]. The non-thermal theory postulates the initiation begins as a result of molecular deformation, but is not consistent with the experiments fact that there is no connection between explosion position and fall given weight. In thermal theory, initiation is triggered by the inhomogeneous heating in the charge. The local hot spot formation and its rapid chemical reaction trigger the initiation. The hot spot theory has been verified by the experiment and observed the hot spot.

Despite the above achievements in theory of initiation and prediction of impact sensitivity, the probabilistic nature of initiation by impact is still elusive. In this letter we will apply the two states model widely used in many areas to formulate a theory to explore the mystery of the probabilistic nature in impact sensitivity.

The two states model has been widely used in many areas. Sometimes the considered system actually has only two sates like a myosin molecule walking on an actin filament[18]: a myosin leg attached to the actin tracks state and the detached state, the two states correspond to the enthalpy dominated and entropy dominated state respectively. In other cases the systems actually have many states, to concentrate on the main factor and easily analysis, we can decompose the many states of the system to two states we concern like the single molecule pull test by mechanical force in DNA and RNA[19-20]. There are many intermediate states between fully folding and unfolding states, to get an clear physical picture and easily cope with, we can decompose so many states to only two state correspond two states: folding and unfolding. To use the two states idea to the initiation, we will decompose the explosive charge state into static and initiation.

To implement the two states models to initiation of explosives by impact, in general, explosives are in metastable states, some activated molecules is exist even in its static state, these molecules will decompose later. As debated above hot spot can give a good explanation of the mechanism of initiation by impact, based hot spot theory and idea of the two states, we will give out our two states model about initiation of explosives by impact. Before given our two states theory, we firstly debate the rationality of hot spot theory. To initiate explosion, it needs enough molecules to decompose to compensate the energy loss by thermal conduct, so the hot spot formation is necessary. On the other hand as the hot spot theory propose the local heating by the impact energy leading the inelastic deformation reasons for the heating[17]; however, as the experiment show the inelastic deformation energy of explosive charge has no relation with the impact energy. In that we propose the hot formation is attributed to the previous existed activated molecules decomposition and the dropped weight initiate the activated. There are no enough activated molecules in static state; as the experiment shown the interaction time scale of dropping weight and explosive charge is about $10^{-4}$ sec, and the explosive charge decomposition time scale is not more than $10^{-6}$ sec[17]. So it exists enough time for the given weight to initiate more activated molecules when the first activated molecules decompose to heat more molecules to activated molecules. Based on above discussion, we propose the number of activated molecules can be initiated is proportional to impact energy.

The hot spot size is about $10^{-4}$cm. As the local heating is a non-equilibrium process, for simplicity we will use an equilibrium effective approach to consider this problem and assume the heating is homogeneous, all the charge molecules surrounding initiated activated molecules are in the same temperature; the probability of explosion is determine by the probability to form hot spot.

Based on above assumptions, let's consider an explosive charge with a volume *V* and *N* molecules, the hot spot critical size is $V_c$, and with $n_c$ active molecules under the given conditions. As in explosive charge molecules are always freeze on lattice, molecules can't move freely to form a hot spot, so the probability to form a hot spot is proportional to the density of active molecules. The probability to form a hot spot is

$$P_{ex} = \rho_{V_c} \cdot \rho_{ac} \tag{1}$$

where $P_{ex}$ is the probability to form a hot spot or initiation, $\rho_{ac}$ is the density of activated molecules density or named the probability in energy space; $\rho_{v_c}$ is a constant for certain sample under given conditions, it can be recognized as the position probability to form a $V_c$ sized hot spot, in that activated molecules must assemble in a $V_c$ sized space, which results from the heat conduct and other inherent properties of explosive charges. For a given sample, if $N \cdot \rho_{ac} < n_c$, the charge can't form a hot spot and the sample is static, otherwise it get some chance to initiation. As debated above, molecules are always not free to move, energy will spread in whole charge. Combined with the asymmetric drop of given weight active molecules are not likely to distribute homogeneously in whole charge. We assume the inhomogeneous distribution lead $\rho_{v_c}$ determine a critical active probability $\rho_{cr}$ for $\rho_{ac}$, which mean if $\rho_{ac} \geq \rho_{cr}$ the explosive will trigger initiation definitely, otherwise it will be probabilistic. For the case $N\rho_{ac} \geq n_c$ the probability of initiation is

$$P_{ex} = \begin{cases} \dfrac{\rho_{ac}}{\rho_{cr}}, & \rho_{ac} < \rho_{cr} \\ 1, & \rho_{ac} \geq \rho_{cr} \end{cases} \tag{2}$$

The density of activated molecule in energy space determine the initiation, to calculate the density, let's consider the partition function of the sample

$$Z = \int_\Omega e^{-\beta E} d\Omega = \int_{E \geq E_{ac}} e^{-\beta E} d\Omega + \int_{E < E_{ac}} e^{-\beta E} d\Omega = Z_{ac} + Z_{in} \tag{3}$$

where $Z$ is the partition function of whole systems, $E_{ac}$ is the activation energy of molecules, $Z_{ac} = \int_{E \geq E_{ac}} e^{-\beta E} d\Omega$ is the part of partition function of activated molecules and $Z_{in} = \int_{E < E_{ac}} e^{-\beta E} d\Omega$ is the inactivated part. Following the same method as two states model, we have decomposed the system molecules state into two parts corresponding two states initiation and static. For simplicity we will use an effective description and express the partition function as

$$Z = g_{ac}e^{-\beta E_{ac}} + g_{in}e^{-\beta E_{in}} \quad (4)$$

where $g_{ac}$ and $g_{in}$ are the density of states of active molecules and inactive molecules respectively, $E_{ac}$ and $E_{in}$ are the effective energy of active and inactive molecules respectively. To calculate the probability of initiation, we need to calculate the distribution of active molecules, and which is

$$\rho_{ac} = \frac{Z_{ac}}{Z} = \frac{1}{1+e^{-\beta \Delta F}} \quad (5)$$

where $\Delta F = \Delta E - T\Delta S$ is the free energy difference, $\Delta E = E_{ac} - E_{in}$ is the energy difference and $\Delta S = k_B \ln(g_{in}/g_{ac})$ is the entropy difference between activated and inactivated molecules. Concluding above discussions, the probability of explosion for the case $N\rho_{ac} \geq n_c$ is

$$p_{ex} = \begin{cases} \frac{1}{\rho_{cr}(1+e^{-\beta \Delta F})}, & \rho_{ac} < \rho_{cr} \\ 1, & \rho_{ac} \geq \rho_{cr} \end{cases} \quad (6)$$

As the given weight dropped onto the explosive charge, molecules will be heated. The heating is inhomogeneous in hot spot theory, for discussion simplicity we assume the heating is homogeneous and the temperature increase is proportional to impact energy. So the temperature after impact is

$$T = T_0 + \alpha W \quad (7)$$

where $T$ is the temperature after impact, $T_0$ is the initial temperature and we will choose room temperature $T_0 = 298K$, $W$ is impact energy, $\alpha$ is a constant for certain explosive under given condition which contains a factor of the decomposition enthalpy of explosive.

To test our model, we compare our theory prediction with RDX experiment impact sensitivity testing data. To test the rationality, we fitted the eq.6 with the experiment data, the results is shown in Figure 1. As Fig.1 shown the theory prediction is good agreement with the experiment results. The fitted activation free energy is 3641.83K (30KJ/mol), which is consistent with experimental thermal decomposition results 40KJ/mol. The fitted $\alpha = 1105.58$, and if all the decomposition enthalpy using to heat the reactive products which will heat to 4000K, the smaller $\alpha$ shows the heat is also heat other surrounded molecules. The position fact $\rho_{cr} = 0.417$.

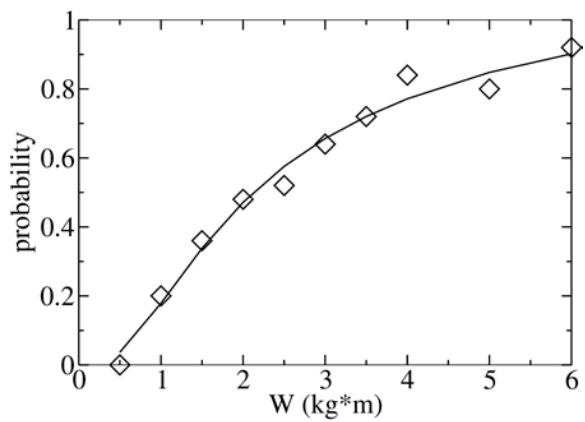

Figure 1. The comparison of theory and RDX experimental impact sensitivity testing data.

In this letter, we propose that the dropped weight initial the existed activated molecules decomposition to initial latter more inactivated molecules to active, which will decompose to initial explosion. Base on the proportionality of impact energy and the initial decomposition molecules, we propose a statistical theory to explain the probabilistic character of initiation of explosives by impact. The theory prediction is shown a good agreement with RDX experimental impact sensitivity data.

Although the above consistence of theory and experimental data, it still need more experiment to verify above assumptions. As our proposition is very different form previous hot spot theory, and all previous experiments are misleading in these two cases. In our proposition the hot spot formation is results from molecules decomposition but dropped weight heating in hot spot theory. In all, the hot spot formation is necessity in our proposition and previous hot spot theory but the formation condition is different.

**Acknowledgements**

The author Gan thanks Mathieu (CEA) for suggestions.

**Competing financial interests**

The author declares no competing financial interests.